\documentclass[a4paper,11pt]{article}
\usepackage{jheppub}
\usepackage{amssymb,amsmath}
\usepackage[normalem]{ulem}
\usepackage[utf8x]{inputenc}
\usepackage{slashed}
\usepackage{graphicx}
\usepackage{csquotes} 
\usepackage{comment}
\usepackage{mathrsfs}
\usepackage{float}
\usepackage{xcolor}
\usepackage{tikz} 
\usepackage{tikz-feynman}
\tikzfeynmanset{compat=1.1.0}
\usepackage[english]{babel}
\usepackage{newlfont}
\usepackage{braket}
\usepackage{epigraph}
\usepackage{graphicx} 
\usepackage{pgfplots}
\usepackage{color}
\usepackage{titlesec}
\usepackage[T1]{fontenc}

\numberwithin{equation}{section}
\newcommand{\vo}{\mathcal{V}}

\preprint{
\begin{minipage}{5cm}
\small
\flushright
KYUSHU-HET-265\\
HU-EP-23/44-RTG
\end{minipage}}

\title{Sequestered String Models imply Split Supersymmetry}

\author{Michele Cicoli$^{1,2}$,} 
\author{Alessandro Cotellucci$^{3}$, and} 
\author{Hajime Otsuka$^{4}$} 
\affiliation{\footnotesize
$^1$ Dipartimento di Fisica e Astronomia, Università di Bologna, via Irnerio 46, 40126 Bologna, Italy\\
$^2$ INFN, Sezione di Bologna, viale Berti Pichat 6/2, 40127 Bologna, Italy\\
$^3$ Humboldt Universität zu Berlin, Institut für Physik and IRIS Adlershof, Zum Grossen Windkanal 6, 12489 Berlin, Germany\\
$^4$ Department of Physics, Kyushu University, 744 Motooka, Nishi-ku, Fukuoka 819-0395, Japan\\
}
\emailAdd{michele.cicoli@unibo.it}
\emailAdd{alessandro.cotellucci@physik.hu-berlin.de}
\emailAdd{otsuka.hajime@phys.kyushu-u.ac.jp}

\abstract{Sequestering is a promising mechanism in 4D string models to reconcile high-scale inflation with low-energy supersymmetry. In this scenario the MSSM lives on branes at singularities and it is sequestered from the sources of supersymmetry breaking in the bulk. The soft-terms are suppressed with respect to the gravitino mass so that all moduli are heavy enough to avoid any cosmological moduli problem. In this paper we study stability bounds and flavour constraints on sequestered string models, finding that they can be satisfied if the soft-terms give rise to a mass spectrum typical of split supersymmetry with TeV-scale gauginos and sfermions around $10^7$ GeV. When instead scalar and gaugino masses are of the same order of magnitude, large flavour changing neutral currents can be avoided only by pushing the soft-terms above $10^6$ GeV. However this scenario is in tension with stability bounds due to the presence of charge and colour breaking vacua which could be populated in the early universe, and the possible emergence of directions along which the potential is unbounded from below.}

\makeatletter
\gdef\@fpheader{}
\makeatother

\begin{document}

\maketitle

\section{Introduction}

One of the main challenges of string phenomenology is to find a 4D string model which can explain all the main hierarchies we see in Nature. Two very important energy scales are the electroweak scale and the Hubble constant during inflation. All of them depend crucially on moduli stabilisation since they are functions of the vacuum expectation values of the string moduli. In particular, a ubiquitous modulus is the one controlling the size of the extra dimensions, i.e. the dimensionless volume $\vo$ of the internal Calabi-Yau space in units of the string length $\ell_s=2\pi\sqrt{\alpha'}$.

A very promising moduli stabilisation mechanism is the Large Volume Scenario (LVS) \cite{Balasubramanian:2005zx,Cicoli:2008va} where the extra dimensional volume is fixed at exponentially large values in terms of the string coupling $g_s\ll 1$ as $\vo\sim e^{c/g_s}\gg 1$ (with $c\sim\mathcal{O}(1)$). In these string models supersymmetry is broken spontaneously by non-zero F-terms of the moduli. The largest contribution comes from the F-term of the volume mode whose fermionic partner is eaten up by the gravitino. The resulting gravitino mass scales as $m_{3/2}\sim M_p/\vo$ and it is naturally much smaller than the Planck mass due to the exponential suppression in $\vo$. In turn, gravitational interactions between the moduli and the Minimal Supersymmetric Standard Model (MSSM) 
mediate supersymmetry breaking to the visible sector, generating non-zero soft-terms. When the MSSM is realised with intersecting and magnetised D7-branes, the soft-terms scale as the gravitino mass: $M_{\mathrm{soft}}\sim m_{3/2}$ \cite{Conlon:2005ki,Conlon:2006wz}. On the other hand, when the MSSM lives on D3-branes at singularities, the visible sector can be sequestered from supersymmetry breaking, resulting in soft-terms which are suppressed with respect to the gravitino mass: $M_{\mathrm{soft}}\ll m_{3/2}$ \cite{Blumenhagen:2009gk,Aparicio:2014wxa,Reece:2015qbf}. 

The last scenario is particularly interesting to reconcile a high inflationary scale with low-energy supersymmetry which is still one of the best solutions to explain the smallness of the electroweak scale. Moreover it avoids any cosmological moduli problem associated with the presence of moduli with masses below $\mathcal{O}(50)$ TeV that would decay after Big Bang Nucleosynthesis \cite{Coughlan:1983ci,Banks:1993en,deCarlos:1993wie}. However a viable model of supersymmetry breaking should also avoid the flavour supersymmetric problem associated to the emergence of dangerously large Flavour Changing Neutral Currents (FCNCs) \cite{Bertolini:1990if,Gabbiani:1996hi,Isidori:2010kg,Misiak:1997ei,Abe:2011rg}, as well as charge and colour breaking (CCB) minima or an unbounded from below (UFB) scalar potential \cite{Casas:1996de,Casas:1995pd}. 

As discussed in \cite{Conlon:2007dw}, these flavour problems can be absent in a so-called mirror mediation scenario where the hidden sector factorises into two parts with no mixing at leading order: the first responsible to determine the flavour structure, and the second which breaks supersymmetry. This framework has been argued to be naturally realised in type IIB LVS models with the MSSM on D7-branes since the flavour structure is determined by the complex structure moduli $U$, while supersymmetry is broken by the K\"ahler moduli $T$. 

In this paper we will instead analyse flavour constraints for sequestered string models where the MSSM lives on D3-branes at singularities. In these cases slepton and squark masses can be generated by the F-terms of either the $T$- or the $U$-moduli depending on the moduli-dependence of the K\"ahler metric for matter fields and the uplifting mechanism to a dS vacuum \cite{Aparicio:2014wxa}. In the first case, mirror mediation is implemented and the mass spectrum is typical of split supersymmetry with the universal gaugino mass $M_{1/2}$ suppressed with respect to the scalar mass $m_0$: $M_{1/2} \sim M_p / \vo^2\ll m_0\sim M_p / \vo^{3/2}\ll m_{3/2} \sim M_p/\vo$. In this scenario TeV-scale gauginos correlate with sfermions around $10^7$ GeV.

On the other hand, when scalar masses are generated by the F-terms of the $U$-moduli, gaugino and scalar masses are of the same order of magnitude, $M_{\mathrm{soft}}\sim M_{1/2}\sim m_0 \sim M_p/\vo^2$, resulting in an MSSM-like mass spectrum. However the same moduli which determine the flavour structure are also responsible to break supersymmetry. This scenario is therefore ruled out by FCNC constraints, unless the overall scale of the soft-terms is raised to very large values, $M_{\mathrm{soft}}\gtrsim 10^6$ GeV, which could be compatible with the observed value of the Higgs mass for $\tan\beta\simeq 2$ \cite{Bagnaschi:2014rsa}. Nonetheless, this scenario is plugged by two stability problems: ($i$) a tree-level scalar potential which is unbounded from below, and ($ii$) the presence of charge and colour breaking vacua. The first problem could be avoided if UFB directions could be lifted by radiative corrections or Planck-suppressed operators, but this detailed analysis is beyond the scope of our paper. On the other hand, the presence of CCB vacua could be harmless if we live instead in a metastable Standard Model-like vacuum with a lifetime longer than the age of the Universe. As pointed out in \cite{Kusenko:1996jn}, this could indeed be the case when $M_{\mathrm{soft}}$ is large, as required to avoid FCNC constraints, since high scale supersymmetry makes CCB minima move away from the SM-like one, making the barrier thicker. Nevertheless, even in this case, it is unclear why CCB minima would not be populated in the early universe during cosmological evolution.

We therefore conclude that sequestered string models are in tension with stability bounds when gauginos and sfermions have comparable masses, while they seem to satisfy current flavour bounds in the case of split supersymmetry.

This article is organised as follows. In Sec.~\ref{sec:Sequestered Supersymmetry Breaking} we review string models with sequestered supersymmetry breaking and their different phenomenological scenarios \cite{Blumenhagen:2009gk,Aparicio:2014wxa,Reece:2015qbf}. In Sec. \ref{sec:Supersymmetric Flavour Problem} we first describe the flavour supersymmetric problem and the conditions for a supersymmetric theory to avoid it, and then we analyse flavour conditions on sequestered string models finding that only split supersymmetry scenarios can be compatible with observations. We finally present our conclusions in Sec. \ref{Concl}.

\section{Sequestering in string models}
\label{sec:Sequestered Supersymmetry Breaking}

In string compactifications sequestering can occur when the MSSM fields are localised in the extra dimensions, like in the case where matter fields are realised as open strings attached to D3-branes at a Calabi-Yau singularity (typically obtained by shrinking a local del Pezzo divisor). In this section we shall describe the main features of sequestered LVS models where moduli stabilisation and supersymmetry breaking are under computational control.

\subsection{The low-energy theory}

For concreteness in what follows we shall focus on the sequestered model analysed in \cite{Aparicio:2014wxa} within the framework of type IIB flux compactifications with O3/O7-planes. The dilaton $S$ (with real part $s$ setting the string coupling $\langle s\rangle=g_s^{-1}$) and the complex structure moduli (collectively denoted as $U$) are fixed at leading order by 3-form fluxes. The K\"ahler moduli are instead stabilised as in standard LVS models. The largest F-terms which break supersymmetry spontaneously are those of the K\"ahler moduli, while the F-terms of the dilaton and the complex structure moduli are subdominant since their stabilisation is at leading order supersymmetric. The MSSM is realised on fractional D3-branes at a del Pezzo singularity (see \cite{Cicoli:2012vw,Cicoli:2013mpa,Cicoli:2013cha,Cicoli:2017shd,Cicoli:2021dhg} for concrete CY orientifold constructions of MSSM-like D3-branes at singularities). The relevant K\"ahler moduli are:
\begin{equation}
T_b=\tau_b+i\theta_b\,, \quad T_s=\tau_s+i\theta_s\,, \quad T_{\mathrm{SM}}=\tau_{\mathrm{SM}}+i\theta_{\mathrm{SM}}\,, \quad G=b+ic\,, 
\end{equation}
where $\tau_b$ and $\tau_s$ are the \emph{big} and \emph{small} LVS divisor volumes with the associated $C_4$-axions $\theta_b$ and $\theta_s$, $\tau_{\mathrm{SM}}\to 0$ is the blow-up mode resolving the MSSM singularity (with corresponding axion $\theta_{\mathrm{SM}}$), and $G$ is the orientifold-odd modulus where $b$ and $c$ arise from the reduction of $B_2$ and $C_2$ on the 2-cycles dual to the shrinking ones.

Several mechanisms can be responsible for uplifting to a dS vacuum. Importantly, as shown in \cite{Aparicio:2014wxa}, they can affect the resulting spectrum of soft-terms as follows (taking three illustrative examples):
\begin{enumerate}
\item Non-zero F-terms of hidden matter fields (T-brane uplifting) \cite{Cicoli:2015ylx}: this case can allow just for a split supersymmetry spectrum since $M_{1/2}\ll m_0\sim m_{\tau_b}\ll m_{3/2}$.

\item Non-zero F-terms of the complex structure moduli \cite{Gallego:2017dvd}: this case can allow just for an MSSM-like spectrum with soft-terms above $\mathcal{O}(50)$ TeV (to avoid a cosmological problem for $\tau_b$) since $M_{1/2}\sim m_0 \sim m_{\tau_b}\ll m_{3/2}$.

\item Non-zero F-term of the blow-up mode of a singularity supporting E(-1)-instantons \cite{Cicoli:2012fh}: depending on the moduli dependence of the matter K\"ahler metric, this case can allow for both split supersymmetry and MSSM-like spectra with TeV-scale gauginos.
\end{enumerate}

Given that only the third case can in principle be compatible with different soft-term spectra, in what follows we shall focus on dS from dilaton-dependent non-perturbative effects even if our results will apply more generically to any uplifting mechanism. For concreteness, we therefore add the K\"ahler modulus $T_{\mathrm{dS}}=\tau_{\mathrm{dS}}+i\theta_{\mathrm{dS}}$ where E(-1)-instantons are localised at $\tau_{\mathrm{dS}}\to 0$.

The $N=1$ low-energy supergravity theory is described by the superpotential:
\begin{equation}
W=W_{\mathrm{flux}}(U,S)+A_s(U,S)\,e^{-a_sT_s}+A_{\mathrm{dS}}(U,S)\,e^{-a_{\mathrm{dS}}(S+\kappa_{\mathrm{dS}}T_{\mathrm{dS}})}+W_{\mathrm{matter}}\,,
\label{eq:W}
\end{equation}
where $W_{\mathrm{flux}}$ is generated by 3-form fluxes \cite{Gukov:1999ya}, the second term is induced by ED3-instantons or gaugino condensation on the small blow-up cycle, the third term is responsible for dS uplifting and originates from E(-1)-instantons, while the matter superpotential reads:
\begin{equation}
W_{\mathrm{matter}}=\mu(U) H_uH_d+\frac{1}{6}Y_{\alpha\beta\gamma}(U)\,C^{\alpha}C^{\beta}C^{\gamma}+...
\end{equation}
where $C^{\alpha}$'s are MSSM superfields and the dots refer to higher dimensional operators. Note that $\mu$ and the Yukawa couplings $Y_{\alpha\beta\gamma}$ can only depend on $U$ due to the axionic shift symmetry of $S$ and $T$ and the fact that $W$ has to be holomorphic. However, explicit computations \cite{Grana:2003ek} have shown that $\mu=0$. Moreover the Yukawa couplings are proportional to the Calabi-Yau holomorphic (3,0)-form for D3-branes at a smooth point, while they are just numbers for D3-branes at singularities. In what follows, we shall therefore set $\mu=0$ but we shall keep a generic dependence of $Y_{\alpha\beta\gamma}$ on $U$ to allow for a potential dependence on the complex structure moduli close to the singularity. The Kähler potential is instead:
\begin{equation}
K=K_{\mathrm{cs}}(U, \bar{U})-\ln(2s)-2\ln\left(\mathcal{V}+\frac{\xi}{2}\,s^{3/2}\right)+\lambda_{\mathrm{SM}}\frac{\tau_{\mathrm{SM}}^2}{\mathcal{V}}+\lambda_b\frac{b^2}{\mathcal{V}}+\lambda_{\mathrm{dS}}\frac{\tau_{\mathrm{dS}}^2}{\mathcal{V}} +K_{\mathrm{matter}}\,,
\end{equation}
where the first two terms give the tree-level K\"ahler potential for the complex structure moduli and the dilaton, the third term is the standard LVS K\"ahler potential with $\mathcal{V}=\tau_b^{3/2}-\tau_s^{3/2}$ and $\mathcal{O}(\alpha'^3)$ corrections proportional to the constant $\xi$ \cite{Becker:2002nn}, the $\lambda$'s are $\mathcal{O}(1)$ coefficients, and the matter K\"ahler potential looks like:
\begin{equation}
K_{\mathrm{matter}}=\tilde{K}_{\alpha\overline{\beta}}(\Phi,\overline{\Phi})\,C^\alpha\overline{C}^{\overline{\beta}}+\left[Z(\Phi,\overline{\Phi})\,H_uH_d+ \mathrm{h.c.}\right],
\label{KmatterGen}
\end{equation}
where $\Phi$ denotes collectively all the moduli and $Z$ is a moduli-dependent Giudice-Masiero coupling. Explicit computations in toroidal cases \cite{Lust:2004cx,Lust:2004fi}, as well as arguments based on locality of the physical Yukawa couplings \cite{Conlon:2006tj}, imply that the matter K\"ahler metric can in general be expanded as (in the singular limit $\tau_{\mathrm{SM}},b\rightarrow 0$):
\begin{equation}
\tilde{K}_{\alpha\overline{\beta}}=\frac{f_{\alpha\overline{\beta}}(U,\overline{U}, S, \overline{S})}{\mathcal{V}^{2/3}}\left(1-c_s\,\frac{\hat{\xi}}{\mathcal{V}}+c_{\mathrm{dS}}\lambda_{\mathrm{dS}}\frac{\tau_{\mathrm{dS}}^2}{\mathcal{V}}\right),
\label{Kmatter}
\end{equation}
where $f_{\alpha\overline{\beta}}(U,\overline{U}, S, \overline{S})$ is an unspecified real function of $U$ and $S$, $\hat\xi\equiv \xi s^{3/2}$, and $c_s$ and $c_{\mathrm{dS}}$ are two coefficients parametrising respectively $\mathcal{O}(\alpha'^3)$ corrections to $\tilde{K}_{\alpha\overline{\beta}}$ and its dependence on the dS modulus. The gauge kinetic function for D3-branes at singularities is:
\begin{equation}
\mathfrak{f}_a=\delta_a S+\kappa_a T_{\mathrm{SM}}\,,
\end{equation}
where $\kappa_a$ is a non-universal numerical coefficient, while $\delta_a$ is a constant which is universal for $\mathbb{Z}_n$ singularities and can be non-universal for more general classes of singularities.

\subsection{Moduli stabilisation}

\subsubsection{D-term fixing}
\label{subsubsec:D-term Stabilisation}

The two moduli $T_{\mathrm{SM}}$ and $G$ are charged under the two anomalous $U(1)$'s of the visible sector with charges $q_1$ and $q_2$. Consequently, the corresponding axions $\theta_{\mathrm{SM}}$ and $c$ get eaten up by the massive Abelian gauge bosons, while $\tau_{\mathrm{SM}}$ and $b$ give rise to moduli-dependent Fayet-Iliopoulos (FI) terms (see \cite{Cicoli:2012vw} for more details). The resulting D-term potential for canonically normalised charged matter fields $\tilde{C}_\alpha$ looks like:
\begin{equation}
V_D=\frac{1}{2\mathrm{Re}(\mathfrak{f}_1)}\left(\sum_{\alpha}q_{1\alpha} |\tilde{C}|^{\alpha}-\xi_1\right)^2+\frac{1}{2\mathrm{Re}(\mathfrak{f}_2)}\left(\sum_{\alpha}q_{2\alpha}|\tilde{C}|^{\alpha}-\xi_2\right)^2,
\end{equation}
where $\mathfrak{f}_1$ and $\mathfrak{f}_2$ are the gauge kinetic functions of the two $U(1)$'s, while the FI-terms are:
\begin{equation}
\begin{aligned}
&\xi_1=-\frac{q_1}{4\pi}\frac{\partial K}{\partial T_{\mathrm{SM}}}=-\frac{q_1\lambda_{\mathrm{SM}}}{4\pi}\frac{\tau_{\mathrm{SM}}}{\mathcal{V}}\,,\\
&\xi_1=-\frac{q_1}{4\pi}\frac{\partial K}{\partial G}=-\frac{q_1\lambda_{b}}{4\pi}\frac{b}{\mathcal{V}}\,.\\
\end{aligned}
\end{equation}
As can be seen from the FI-terms, the D-term potential is a leading order effect since it scales with the exponentially large volume as $V_D\sim \mathcal{V}^{-2}$. D-term stabilisation has therefore to be supersymmetric at leading order. Vanishing vacuum expectation values of the visible matter fields then imply $\xi_1=\xi_2=0$, forcing $\tau_{\mathrm{SM}}$ and $b$ to the singular regime, i.e. $\langle\tau_{\mathrm{SM}}\rangle=\langle b\rangle=0$. Similarly, also the $T_{\mathrm{dS}}$ modulus can be fixed in the singular regime by minimising the hidden sector D-term potential (focusing for simplicity on canonically normalised hidden fields $\phi_{h,i}$ with charges $q_{h,i}$ under an anomalous $U(1)$):
\begin{equation}
V_{D}^{\mathrm{dS}}=\frac{1}{2\mathrm{Re}(\mathfrak{f}_h)}\left(\sum_{i}q_{h,i}|\phi_{h,i}|^2-\xi_h\right)^2,
\label{eq:D-term_dS1}
\end{equation}
with FI-term:
\begin{equation}
\xi_h=-\frac{q_{\mathrm{dS}}}{4\pi}\frac{\partial K}{\partial T_{\mathrm{dS}}}=-\frac{q_{\mathrm{dS}}}{4\pi}\frac{\lambda_{\mathrm{dS}}\tau_{\mathrm{dS}}}{\mathcal{V}}.
\label{eq:D-term_dS2}
\end{equation}
If the hidden matter fields $\phi_{h,i}$ are fixed by the corresponding F-terms at $\langle\sum_iq_{h,i}|\phi_{h,i}|^2\rangle=0$, the D-term potential vanishes for $\xi_h=0$ implying $\langle\tau_{\mathrm{dS}}\rangle=0$ which is compatible with an E(-1)-instanton at a singularity.

\subsubsection{F-term fixing}
\label{subsubsec:F-term Stabilisation}

It is straightforward to realise that the flux-generated scalar potential scales as the D-term potential: $V_{\mathrm{flux}}\sim \mathcal{V}^{-2}$. Therefore its minimisation has to be again supersymmetric. Setting to zero the F-terms of the dilaton and the complex structure moduli corresponds to fixing these fields at:
\begin{equation}
D_S W_{\mathrm{flux}}|_{\xi=0}=0\qquad\text{and} \qquad  D_U W_{\mathrm{flux}}|_{\xi=0}=0\,.
\label{FSFU=0}
\end{equation}
Substituting the stabilised values of $S$ and $U$ into $W_{\mathrm{flux}}$, the flux superpotential can be considered as constant: $\langle W_{\mathrm{flux}}\rangle\equiv W_0$. The remaining K\"ahler moduli are lifted at subleading order thanks to $\alpha'$ and non-perturbative corrections which break the no-scale structure generating a standard LVS potential of the form \cite{Balasubramanian:2005zx,Cicoli:2008va} (after fixing the axion $\theta_s$ at $e^{-a_s\theta_s}=-1$ and setting $e^{K_{\mathrm{cs}}}=1$):
\begin{equation}
V_F^{\mathrm{LVS}}=\frac{1}{2s}\left[\frac{8}{3}\frac{A_s^2a_s^2e^{-2a_s\tau_s}\sqrt{\tau_s}}{\mathcal{V}} -4\frac{W_0A_sa_se^{-a_s\tau_s}\tau_s}{\mathcal{V}^2}+\frac{3\hat{\xi} W_0^2}{4\mathcal{V}^3}\right].
\label{VLVS}
\end{equation}
This potential admits a supersymmetry breaking AdS minimum located at (for $a_s\tau_s\gg 1$):
\begin{align}
\langle\tau_s\rangle^{3/2}&\simeq \frac{\hat{\xi}}{2}\qquad\text{and}
\qquad
\langle\mathcal{V}\rangle \simeq \frac{3W_0\sqrt{\langle\tau_s\rangle}}{4a_sA_s}e^{a_s\langle\tau_s\rangle}\,.
\label{eq:VF}
\end{align}
Given that (\ref{VLVS}) depends on $S$ and $U$ (via $A_s=A_s(S,U)$ and $\hat\xi=\xi s^{3/2}$), the LVS potential causes a shift of the vacuum expectation values for the dilaton and the complex structure moduli determined by the supersymmetric condition (\ref{FSFU=0}). These shifts can be parametrised in terms of the functions $\omega_S(U,S)$ and $\omega_U (U,S)$ as follows:
\begin{eqnarray}
D_S W &\simeq& D_S W_{\mathrm{flux}}|_{\xi=0}-\frac{3\omega_S(U,S)\hat{\xi}W_0}{4s\mathcal{V}}=-\frac{3\omega_S(U,S)\hat{\xi}W_0}{4s\mathcal{V}}\,, \\
D_U W &\simeq& D_U W_{\mathrm{flux}}|_{\xi=0}-\frac{3\omega_U (U,S)\hat{\xi}W_0}{4s\mathcal{V}}=-\frac{3\omega_U(U,S)\hat{\xi}W_0}{4s\mathcal{V}}\,.
\end{eqnarray}
As pointed out in \cite{Blumenhagen:2009gk}, these shifts induce non-zero F-terms for $S$ and $U$ which affect the soft-terms significantly, as we will show later.

Following \cite{Cicoli:2012fh}, a dS vacuum can be realised by E(-1)-instanton corrections to the superpotential (\ref{eq:W}) which induce the following F-term contribution (after fixing the axion $\theta_{\mathrm{dS}}$ at $e^{-i2a_{\mathrm{dS}}\theta_{\mathrm{dS}}}=-1$):
\begin{equation}
V_F^{\mathrm{dS}}=\frac{(\kappa_{\mathrm{dS}}a_{\mathrm{dS}}A_{\mathrm{dS}})^2}{s\lambda_{\mathrm{dS}}}\frac{e^{-2a_{\mathrm{dS}}(s+\kappa_{\mathrm{dS}}\tau_{\mathrm{dS}})}}{\mathcal{V}} = \frac{(\kappa_{\mathrm{dS}}a_{\mathrm{dS}}A_{\mathrm{dS}})^2}{s\lambda_{\mathrm{dS}}}\frac{e^{-2a_{\mathrm{dS}}s}}{\mathcal{V}}\,,
\end{equation}
where we have substituted $\langle\tau_{\mathrm{dS}}\rangle=0$ from D-term stabilisation.\footnote{Note that these minimisation relations are exact only at leading order. Subleading corrections give rise however just to negligible effects.} This potential represents an uplifting contribution which has to be added to the LVS potential (\ref{VLVS}). The minimum of the total potential can then be set to zero by tuning the dilaton via an appropriate choice of 3-form fluxes. If this cancellation is required up to ${\cal O}({\cal V}^{-3})$, the dilaton has to be tuned such that
\begin{equation}
e^{-2a_{\mathrm{dS}}s}=\frac{9}{128}\left(\frac{W_0}{\kappa_{\mathrm{dS}}a_{\mathrm{dS}}A_{\mathrm{dS}}}\right)^2\frac{\hat{\xi}\lambda_{\mathrm{dS}}}{a_s\tau_s\mathcal{V}^2}\,.
\end{equation}

\subsubsection{F-terms}
\label{subsubsec:F-terms}

Supersymmetry is spontaneously broken by non-zero F-terms of the moduli (non-zero D-terms also develop due to subleading effects), and so the gravitino becomes massive. Its mass will be used as a reference for all masses in the model and looks like:
\begin{equation}
m_{3/2}=e^{K/2}|W| M_p\simeq\frac{W_0}{\sqrt{2s}}\frac{M_p}{\mathcal{V}}.
\end{equation}
The F-terms of the K\"ahler moduli in the geometric regime are given by:
\begin{eqnarray}
F^{T_b}&\simeq& -\sqrt{\frac{2}{s}}\,\frac{\tau_b W_0}{\mathcal{V}}\sim \mathcal{O}\left(m_{3/2}^{1/3}M_p^{5/3}\right), \\ 
F^{T_s}&\simeq& - \left(\frac{3}{2a_s\sqrt{2s}}\right)\frac{W_0}{\mathcal{V}}\sim \mathcal{O}\left(m_{3/2}M_p\right).
\end{eqnarray}
These are the largest F-terms with the dominant F-term given by $F^{T_b}$, showing that the gravitino becomes massive by eating up the fermionic partner of $T_b$. As argued above, non-zero F-terms for the dilaton and the complex structure moduli are induced by quantum corrections beyond semi-classical level, and read:
\begin{eqnarray}
F^S &\simeq& - \left(\sqrt{\frac{s}{2}}\frac{3\hat\xi}{2}\omega'_S(U,S)\right)\frac{W_0}{\mathcal{V}^2} \sim \mathcal{O}\left(m_{3/2}^2\right), \label{FS} \\
F^U &\simeq& \left(\frac{K^{U\overline{U}}}{2 s^2}\frac{\overline{\omega}_U(U,S)}{\omega_S'(U,S)}\right) F^S \equiv \beta^U (U,S)\,F^S\sim \mathcal{O}(m_{3/2}^2)\,,
\label{FU}
\end{eqnarray}
where $\omega_S'(U,S)\equiv 3+2\overline{\omega}_S(U,S)$ and  $\beta_U(U,S)$ is an $\mathcal{O}(1)$ function of $S$ and $U$ in the volume expansion. Let us stress that these two F-terms are hierarchically smaller than the F-terms of $T_b$ and $T_s$. The F-terms of the moduli localised at the MSSM singularity vanish: 
\begin{equation}
F^{T_{\mathrm{SM}}}=0 \qquad\text{and}\qquad F^G=0\,,
\end{equation}
which is a crucial result to guarantee that the visible sector is sequestered from supersymmetry breaking. Finally the F-term of the dS K\"ahler modulus $T_{\mathrm{dS}}$ scales as:
\begin{equation}
F^{T_{\mathrm{dS}}}\simeq\left(\frac34\sqrt{\frac{\hat\xi}{\lambda_{\mathrm{dS}} s a_s \tau_s}}\right) \frac{W_0}{\mathcal{V}}\sim \mathcal{O}\left(m_{3/2}M_p \right).
\end{equation}

\subsection{Soft supersymmetry breaking terms}
\label{subsec:Soft-Terms}

The soft supersymmetry breaking Lagrangian takes the general form:
\begin{equation}
\mathcal{L}_{\mathrm{soft}}=\frac{1}{2}\left(M_a\hat{\lambda}^a\hat{\lambda}^a+h.c.\right)-m_{\alpha}^2 \hat{C}^\alpha \overline{\hat{C}}^{\overline{\alpha}} -\left(\frac{1}{6}A_{\alpha\beta\gamma}\hat{Y}_{\alpha\beta\gamma}\hat{C}^{\alpha}\hat{C}^{\beta}\hat{C}^{\gamma}+B\hat\mu \hat{H}_u \hat{H}_d+ \mathrm{h.c.}\right),
\label{Lsoft}
\end{equation}
where we have focused for simplicity on a diagonal K\"ahler matter metric, i.e. $\tilde{K}_{\alpha\overline{\beta}} = \tilde{K}_\alpha \delta_{\alpha\overline{\beta}}$, and all fields have been canonically normalised. Before computing the form of the soft-terms, we discuss the moduli dependence of the K\"ahler matter metric which distinguishes two phenomenologically different scenarios.

\subsubsection*{Matter metric}

The K\"ahler matter metric takes the form (\ref{Kmatter}) where $f_{\alpha\overline{\beta}}$, or simply $f_\alpha$ in the diagonal case, is a priori a general real function of the dilaton and the complex structure moduli, while $c_s$ and $c_{\mathrm{dS}}$ are two unspecified parameters. As pointed out in \cite{Conlon:2006tj}, this expression guarantees that the physical Yukawa couplings:
\begin{equation}
\hat{Y}_{\alpha\beta\gamma} = e^{K/2}\,\frac{Y_{\alpha\beta\gamma}}{\sqrt{\tilde{K}_\alpha\tilde{K}_\beta\tilde{K}_\gamma}}\,,   
\end{equation}
do not depend on the overall volume at leading order. A cancellation of the volume dependence at all orders would instead correspond to:
\begin{equation}
\tilde{K}_{\alpha}=h_{\alpha}(U,\overline{U},S,\overline{S})\,e^{K/3}\,,
\label{LocRel}
\end{equation}
where $h_{\alpha}$ is another general real function of $S$ and $U$ related to $f_{\alpha}$ as:
\begin{equation}
f_{\alpha}=\frac{h_{\alpha}\,e^{K_{\mathrm{cs}}/3}}{(2s)^{1/3}}\,.
\end{equation}
Following \cite{Aparicio:2014wxa}, we shall consider two cases:
\begin{itemize}
\item \emph{Local limit}: in this case the relation (\ref{LocRel}) holds only at leading order.
\item \emph{Ultra-local limit}: in this case the relation (\ref{LocRel}) holds also at subleading order, implying $c_s=c_{\mathrm{dS}}=1/3$ by comparing (\ref{Kmatter}) with (\ref{LocRel}).
\end{itemize}
As we will see, the local limit reproduces a split supersymmetry scenario whereas the ultra-local limit is characterised by a mass spectrum where gaugino and scalar masses are of the same order of magnitude. 

\subsubsection{Gaugino masses}
\label{subsubsection:Gaugino Masses}

The general expression of gaugino masses in gravity mediated supersymmetry breaking is:
\begin{equation}
M_a=\frac{1}{2}\frac{F^m\partial_m \mathfrak {f}_a}{\text{Re}(\mathfrak{f}_a)}\,.
\end{equation}
Given that in our case $\mathfrak {f}_a=\delta_a S+\kappa_a T_{\mathrm{SM}}$ and $F^{T_{\mathrm{SM}}}=0$, the only non-zero contribution comes from the F-term of the dilaton. Focusing on universal $\delta_a$'s, the resulting gaugino masses are hierarchically smaller than the gravitino mass:
\begin{equation}
M_{1/2}=\frac{F^S}{2s}=
\left(\frac{3\hat\xi}{4\sqrt{2s}}\omega'_S\right)\frac{W_0}{\mathcal{V}^2} \sim \mathcal{O}\left(\frac{m_{3/2}^2}{M_p}\right).
\end{equation}

\subsubsection{Scalar masses}
\label{subsubsec:Scalar Masses}

Slepton and squark masses receive contributions from both F- and D-terms which we will discuss separately. 

\subsubsection*{F-term contribution}

The F-term contribution to the scalar masses for a diagonal K\"ahler matter metric is:
\begin{equation}
m_{\alpha}^2|_F=m_{3/2}^2-\overline{F}^{\overline{m}}F^n\partial_{\overline{m}}\partial_n \ln \tilde{K}_{\alpha}\,.
\label{GenFScalMass}
\end{equation}
Let us consider the local and ultra-local scenarios separately.

\subsubsection*{Local limit}

To obtain the leading order F-term contribution to scalar masses in the local scenario, we can rewrite the K\"ahler matter metric (\ref{Kmatter}) as:
\begin{equation}
\tilde{K}_\alpha \simeq h_\alpha\,e^{K/3}\left(\frac{1-c_s\frac{\hat\xi}{\mathcal{V}}}{1-\frac{\hat\xi}{3\mathcal{V}}}\right).
\label{Klocal}
\end{equation}
Hence the general expression (\ref{GenFScalMass}) becomes:
\begin{eqnarray}
m_{\alpha}^2&\simeq & m_{3/2}^2-\overline{F}^{\overline{m}}F^n\partial_{\overline{m}}\partial_n\left[\frac{1}{3}K+\ln\left(\frac{1-c_s\frac{\hat{\xi}}{\mathcal{V}}}{1-\frac{\hat{\xi}}{3\mathcal{V}}}\right)+\ln h_{\alpha}\right] \nonumber \\
&\simeq &-\overline{F}^{\overline{m}}F^n\partial_{\overline{m}}\partial_n\left[ \ln\left(\frac{1-c_s\frac{\hat{\xi}}{\mathcal{V}}}{1-\frac{\hat{\xi}}{3\mathcal{V}}}\right)+\ln h_{\alpha}\right] \nonumber \\
&\simeq &\frac{15}{16}\left(c_s-\frac{1}{3}\right)\left(\frac{F^{T_b}}{\tau_b}\right)^2\frac{\hat{\xi}}{\mathcal{V}}+\mathcal{O}\left(\frac{1}{\mathcal{V}^4}\right) \nonumber \\
&\simeq &\left(c_s-\frac{1}{3}\right)\frac{5}{\omega'_S}\,m_{3/2}M_{1/2}\sim \mathcal{O}(M_{1/2}m_{3/2})\,.
\label{m0Local}
\end{eqnarray}
It is worth pointing out that in the local scenario squark and scalar masses are mainly generated by the F-term of $T_b$, and so turn out to be flavour universal, i.e. $m_\alpha \simeq m_0$ $\forall \alpha$, and with a hierarchy typical of split supersymmetry:
\begin{equation}
M_{1/2}\sim \frac{M_p}{\mathcal{V}^2}\ll m_0 \sim \frac{M_p}{\mathcal{V}^{3/2}}\ll m_{3/2}\sim \frac{M_p}{\mathcal{V}}\,.
\end{equation}
Note that we need to require $c_s>1/3$ to avoid tachyons. For $c_s=1/3$, the leading result would instead be vanishing, a clear sign of the ultra-local limit which we will study below.

\subsubsection*{Ultra-local limit}

In the ultra-local scenario we shall instead focus on a K\"ahler matter metric of the form $\tilde{K}_\alpha\simeq h_\alpha\, e^{K/3}$ which gives:
\begin{eqnarray}
m_{\alpha}^2 &\simeq & m_{3/2}^2-\overline{F}^{\overline{m}}F^n\partial_{\overline{m}}\partial_n\left(\frac{1}{3}K+\ln h_{\alpha}\right) \nonumber \\
&\simeq & -\overline{F}^{\overline{m}}F^n\partial_{\overline{m}}\partial_n \ln h_{\alpha} \simeq g_\alpha\,\left(\frac{F^S}{2s}\right)^2 
\simeq g_\alpha\,M_{1/2}^2 \sim \mathcal{O}(M_{1/2}^2)\,,
\end{eqnarray}
where $g_\alpha$ is a flavour-dependent function of $S$ and $U$ defined as:
\begin{equation}
g_\alpha(U,\overline{U},S,\overline{S})\equiv -4s^2\left(\partial_{\overline{S}}\partial_S+\beta^U \partial_{\overline{S}}\partial_U+\overline{\beta}^{\overline{U}}\partial_{\overline{U}}\partial_S+\overline{\beta}^{\overline{U}}\beta^U\partial_{\overline{U}}\partial_U\right)\ln h_\alpha\,.
\label{gFunction}
\end{equation}
Contrary to the local limit, in this case the main contribution comes from $F^S$ and $F^U$ which is proportional to $F^S$, as can be seen from (\ref{FU}). Thus scalar masses turn out to be of the same order of gaugino masses and, above all, they are not flavour universal since the function $g_\alpha$ defined in (\ref{gFunction}) is in general different from $h_\alpha$. This can be a potential source of large FCNCs, as we will analyse in Sec. \ref{sec:Supersymmetric Flavour Problem}. 

More in general, in the ultra-local case the assumption of a diagonal K\"ahler matter metric seems inappropriate since the F-terms of the dilaton and the complex structure moduli would in general generate also non-zero off-diagonal entries of the scalar mass matrix. In fact, a general K\"ahler matter metric:
\begin{equation}
\tilde{K}_{\alpha\overline{\beta}} \simeq h_{\alpha\overline{\beta}}\,e^{K/3}\,,   
\end{equation}
would generate soft masses for the unnormalised scalars of the form:
\begin{eqnarray}
\tilde{m}^2_{\alpha\overline{\beta}} &=& m_{3/2}^2 \tilde{K}_{\alpha\overline{\beta}} -\overline{F}^{\overline{m}}F^n\left(\partial_{\overline{m}}\partial_n \tilde{K}_{\alpha\overline{\beta}} -
(\partial_{\overline{m}}\tilde{K}_{\alpha\overline{\gamma}})\tilde{K}^{\overline{\gamma}\delta}(\partial_n\tilde{K}_{\delta\overline{\beta}})\right) \nonumber \\ 
&\simeq& - e^{K/3}\,\sum_{m,n \in\{S,U\}}\overline{F}^{\overline{m}}F^n\left(\partial_{\overline{m}}\partial_n h_{\alpha\overline{\beta}} -
(\partial_{\overline{m}} h_{\alpha\overline{\gamma}}) h^{\overline{\gamma}\delta}(\partial_n h_{\delta\overline{\beta}})\right).
\end{eqnarray}
Canonical normalisation would remove the $e^{K/3}$ factor, inducing the following mass matrix for the normalised squarks and sleptons:
\begin{equation}
m^2_{\alpha\overline{\beta}} = g_{\alpha\overline{\beta}}\,M_{1/2}^2\,,  
\label{m0UltraLoc}
\end{equation}
where $g_{\alpha\overline{\beta}}(U,\overline{U},S,\overline{S})$ is an $\mathcal{O}(1)$ function of $S$ and $U$ which is in general different from $h_{\alpha\overline{\beta}}$, showing that off-diagonal terms are expected to be of the same order of magnitude as the diagonal ones.

\subsubsection*{D-term contribution}

The D-term contribution to scalar masses is given by \cite{Dudas:2005vv}:
\begin{equation}
m_{\alpha}^2|_{D}=\tilde{K}_{\alpha}^{-1}\sum_ig_i^2D_i\partial^2_{\alpha\overline{\alpha}}D_i-V_{D,0}.
\end{equation}
In our case, the relevant D-term is the one associated to the anomalous $U(1)$ symmetry at the singularity which supports the non-perturbative effects responsible for the dS uplift:
\begin{equation}
D_{T_{\mathrm{dS}}}=\frac{q_{\mathrm{dS}}}{4\pi}\partial_{T_{\mathrm{dS}}}K\,.
\end{equation}
Using $g_{T_{\mathrm{dS}}}^2 = s^{-1}$, we find a universal contribution to scalar masses of the form:
\begin{eqnarray}
m_0^2|_D&=&\frac{q_{\mathrm{dS}}}{4\pi}\tilde{K}_{\alpha}^{-1}g_{T_{\mathrm{dS}}}^2D_{T_{\mathrm{dS}}}\partial_{T_{\mathrm{dS}}} \partial^2_{C\overline{C}}K-V_{D,0} \nonumber \\
&=& \frac{q_{\mathrm{dS}}}{4\pi s}\tilde{K}_{\alpha}^{-1}D_{T_{\mathrm{dS}}}\partial_{T_{\mathrm{dS}}}\tilde{K}_{\alpha}-V_{D,0}=(2c_{\mathrm{dS}}-1)V_{D,0}\sim \mathcal{O}(\mathcal{V}^{-4})\,,
\end{eqnarray}
which scales as $V_{D,0} \sim \mathcal{O}(\mathcal{V}^{-4})$, as can be easily seen by considering subdominant corrections to D-term moduli stabilisation. Hence, in the local scenario the D-term contribution is negligible with respect to the F-term one. On the other hand, in the ultra-local scenario D-term soft scalar masses vanish for $c_{\mathrm{dS}}=1/3$. This can be seen by including in the D-term expression also the contribution from the vacuum expectation value of the F-term potential $-V_{F,0}/3=V_{D,0}/3$, giving:\footnote{Strictly speaking, we should have included this term also in the local case but we omitted it since it affects the final result only quantitatively but not qualitatively.}
\begin{equation}
m_0^2=2\left(c_{\mathrm{dS}}-\frac{1}{3}\right)V_{D,0}=0\,.
\end{equation}

\subsubsection{Trilinear terms}
\label{subsubsec:Trilinear terms}

The trilinear A-terms for a diagonal K\"ahler matter metric are given by:
\begin{equation}
A_{\alpha\beta\gamma}=F^m\left[K_m+\partial_m\ln Y_{\alpha\beta\gamma}-\partial_m\ln\left(\tilde{K}_{\alpha}\tilde{K}_{\beta}\tilde{K}_{\gamma}\right)\right].
\end{equation}
Writing again the K\"ahler matter metric as in (\ref{Klocal}), where $c_s>1/3$ in the local limit and $c_s=1/3$ in the ultra-local case, the trilinear A-terms turn out to be:
\begin{eqnarray}
A_{\alpha\beta\gamma}&\simeq& F^m\partial_m\left[K-\ln\left(e^Kh_{\alpha}h_{\beta}h_{\gamma}\right)-3\ln\left(\frac{1-c_s\frac{\hat{\xi}}{\mathcal{V}}}{1-\frac{\hat{\xi}}{3\mathcal{V}}}\right)+\ln Y_{\alpha\beta\gamma}\right] \nonumber \\
&\simeq& F^m\partial_m\left[\ln\left(\frac{ Y_{\alpha\beta\gamma}}{h_{\alpha}h_{\beta}h_{\gamma}}\right)+3\left(c_s-\frac{1}{3}\right)\frac{\hat{\xi}}{\mathcal{V}}\right] \nonumber \\
&\simeq& \left[ y_{\alpha\beta\gamma}+\frac{6}{\omega'_S}\left(c_s-\frac{1}{3}\right)\right] M_{1/2}\sim  \mathcal{O}(M_{1/2})\,,
\label{Aterms}
\end{eqnarray}
where $y_{\alpha\beta\gamma}$ is a flavour-dependent function of $S$ and $U$ defined as:
\begin{equation}
y_{\alpha\beta\gamma}(S,U)\equiv 2 s\left(\partial_S +\beta^U \partial_U\right)\ln\left(\frac{ Y_{\alpha\beta\gamma}}{h_{\alpha}h_{\beta}h_{\gamma}}\right).
\end{equation}
The final result is proportional to the gaugino mass both in the local and in the ultra-local limit where the last term in (\ref{Aterms}) vanishes for $c_s=1/3$. The first term in (\ref{Aterms}) is generated by $F^S$ and $F^U$ and it proportional to $y_{\alpha\beta\gamma}$. Hence, it breaks flavour universality contrary to the second term which is flavour universal since it is generated by $F^{T_b}$. The first contribution is present both in the local and ultra-local limits, while the second one is present only in the local case. However the two contributions are in general of the same order of magnitude, and so we expect that in both limits the physical A-terms $\hat{A}_{\alpha\beta\gamma} = A_{\alpha\beta\gamma} \hat{Y}_{\alpha\beta\gamma}$ are not proportional to the physical Yukawa couplings $\hat{Y}_{\alpha\beta\gamma}$. This is another manifestation of the supersymmetric flavour problem which we will discuss in Sec. \ref{sec:Supersymmetric Flavour Problem}.

\subsubsection{$\hat\mu$ and $B\hat\mu$-terms}
\label{subsubsec:Bmu-term}

Since the MSSM lives at the singularity, we shall assume that the Giudice-Masiero coupling $Z$ in (\ref{KmatterGen}) scales as the K\"ahler matter metric (\ref{Kmatter}) but with the function $f_{\alpha\overline{\beta}}$ replaced by the generic real function of $S$ and $U$ $\gamma(S,\overline{S},U,\overline{U})$. Moreover, we shall focus again for simplicity on a diagonal K\"ahler matter metric. As explained in \cite{Aparicio:2014wxa}, the physical $\hat\mu$ term becomes:
\begin{equation}
\hat\mu = \frac{\gamma}{\sqrt{f_{H_u}f_{H_d}}}\left[\frac{3}{\omega_S'}\left(c_s-\frac13\right)-2s\left(\partial_{\overline{S}}+\overline{\beta}^{\overline{U}}\partial_{\overline{U}}\right)\ln\gamma\right] M_{1/2}\sim\mathcal{O}(M_{1/2})\,,
\end{equation}
and so it is of the same order of magnitude of the gaugino mass both in the local and in the ultra-local limit where $c_s=1/3$.

Interestingly, as shown in \cite{Aparicio:2014wxa}, both F- and D-term contributions to the $B\hat\mu$-term turn out to be proportional to the scalar masses:
\begin{equation}
B\hat\mu|_{F,D}=\frac{\gamma}{\sqrt{f_{H_u}f_{H_d}}}\,m_0^2|_{F,D}\,.
\end{equation}
Thus in the local limit $B\hat\mu\sim\mathcal{O}(M_{1/2}m_{3/2})$, whereas in the ultra-local limit $B\hat\mu\sim\mathcal{O}(M_{1/2}^2)$.

\subsubsection{Mass spectrum}
\label{subsubsec:Mass Spectra}

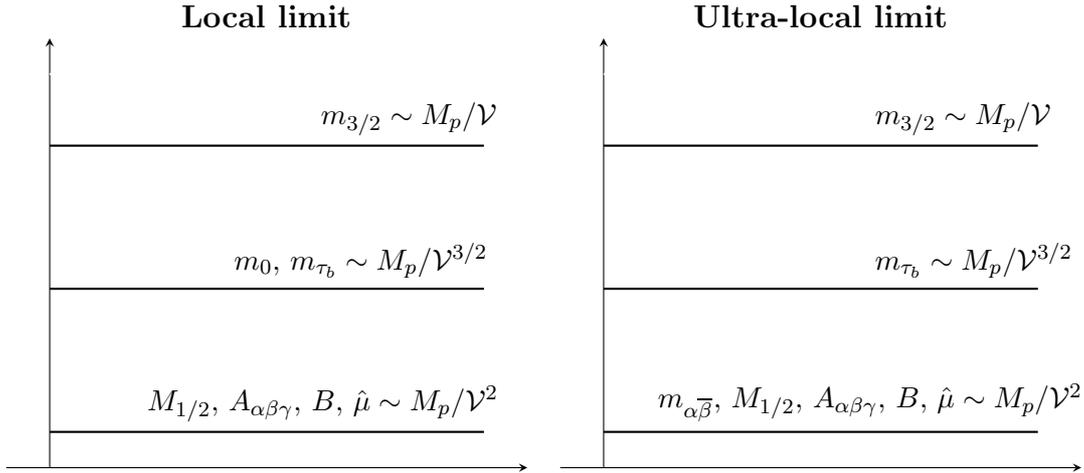
\begin{figure}[ht]
\begin{tabular}[c]{cc}
\begin{tikzpicture}
\begin{axis} [axis lines=middle,
enlargelimits,
xtick={0},ytick={0},
xticklabels={$0$},
yticklabels={$0$},
xlabel=,ylabel=]
\addplot [domain=0:3,
samples=40,smooth,thick,white]
{4}node [pos=0.6, above right] {};
\addplot [domain=0:3,
samples=40,smooth,thick,black]
{3.5}node [pos=0.6, above right] {$m_{3/2}\sim M_p/\mathcal{V}$};
\addplot [domain=0:3,
samples=40,smooth,thick,black]
{2.5}node [pos=0.4, above right] {$m_0,\,m_{\tau_b}\sim M_p/\mathcal{V}^{3/2}$};
\addplot [domain=0:3,
samples=40,smooth,thick,black]
{1.5}node [pos=0.2, above right] {$M_{1/2},\,A_{\alpha\beta\gamma},\,B,\,\hat\mu\sim M_p/\mathcal{V}^2$};
\end{axis}
\node[above,font=\large\bfseries] at (current bounding box.north) {Local limit};
\end{tikzpicture}&
\begin{tikzpicture}
\begin{axis} [axis lines=middle,
enlargelimits,
xtick={0},ytick={0},
xticklabels={$0$},
yticklabels={$0$},
xlabel=,ylabel=]
\addplot [domain=0:3,
samples=40,smooth,thick,white]
{4}node [pos=0.6, above right] {};
\addplot [domain=0:3,
samples=40,smooth,thick,black]
{3.5}node [pos=0.6, above right] {$m_{3/2}\sim M_p/\mathcal{V}$};
\addplot [domain=0:3,
samples=40,smooth,thick,black]
{2.5}node [pos=0.6, above right] {$m_{\tau_b}\sim M_p/\mathcal{V}^{3/2}$};
\addplot [domain=0:3,
samples=40,smooth,thick,black]
{1.5}node [pos=0.1, above right] {$m_{\alpha\overline{\beta}},\,M_{1/2},\,A_{\alpha\beta\gamma},\,B,\,\hat\mu\sim M_p/\mathcal{V}^2$};
\end{axis}
\node[above,font=\large\bfseries] at (current bounding box.north) {Ultra-local limit};
\end{tikzpicture}
\end{tabular}
\caption{Scaling of the soft-terms for the local (left) and ultra-local scenario (right).}
\label{fig:ms}
\end{figure}

The resulting structure of the soft-terms for the local and ultra-local limits are summarised in Fig. \ref{fig:ms}.

\newpage
Let us comment on the two cases separately:
\begin{itemize}
\item The \emph{local limit} is characterised by the following main features:
\begin{enumerate}
\item Squarks and sleptons are hierarchically heavier than gauginos, reproducing a typical split supersymmetry spectrum.

\item Scalar masses are of the same order of the mass of the lightest K\"ahler modulus $\tau_b$.

\item Scalar masses are flavour universal and proportional to the identity.

\item The physical trilinear A-terms scale as $M_{1/2}$ and are in general not proportional to the physical Yukawa couplings.

\item As pointed out in \cite{Aparicio:2014wxa}, TeV-scale gauginos can be obtained for a compactification volume of order $\mathcal{V}\sim 10^7$ which, in turn, gives $m_{3/2}\sim 10^{10}$ GeV and $m_0\sim m_{\tau_b}\sim 10^7$ GeV. The volume modulus $\tau_b$ is therefore heavy enough to avoid any cosmological problem.
\end{enumerate}

\item In the \emph{ultra-local limit} the soft-terms have the following interesting properties:
\begin{enumerate}
\item Scalar and gaugino masses are of the same order of magnitude, yielding a standard MSSM-like spectrum.

\item The lightest K\"ahler modulus $\tau_b$ is hierarchically heavier than the superpartners, and so its mass is well above the lower bound from the cosmological moduli problem.

\item Squark and slepton masses are in general not flavour universal and feature off-diagonal terms of the same order as the diagonal ones.

\item The physical trilinear A-terms are in general not proportional to the physical Yukawa couplings.

\item An MSSM-like spectrum can also be obtained when the visible sector lives on D7-branes \cite{Conlon:2006wz}. This situation has two important differences with the ultra-local D3-brane case due to the absence of sequestering which leads to a \emph{pro} and a \emph{con}. The \emph{pro} is that the soft-terms are mainly generated by the F-terms of the K\"ahler moduli which do not induce any flavour problem. The \emph{con} is that the soft-terms are of order $m_{3/2}$, and so TeV-scale supersymmetry is in tension with high-scale inflation and the cosmological moduli problem for $\tau_b$.
\end{enumerate}
\end{itemize}

\section{Supersymmetric flavour problem}
\label{sec:Supersymmetric Flavour Problem}

As we have already seen, the soft-terms can have a non-trivial flavour structure which sources new flavour mixing contributions that can be in tension with observations. This is the origin of the famous \emph{supersymmetric flavour problem} \cite{Gabbiani:1996hi,Isidori:2010kg,Luty:2005sn,Dimopoulos:1995ju,Misiak:1997ei,Abe:2011rg,deCarlos:1995ah} which in this section we will analyse in sequestered string models.

\subsection{Experimental constraints}
\label{ExpConstr}

In order to avoid the supersymmetric flavour problem, the new sources of flavour violation from supersymmetry breaking should be small. This is guaranteed if the scalar masses are nearly flavour universal and the trilinear A-terms are dominantly proportional to the physical Yukawa couplings. If the relevant soft-terms are written as:
\begin{equation}
m^2_{\alpha\overline{\beta}}=m_0^2\,\delta_{\alpha\overline{\beta}} +\Delta m^2_{\alpha\overline{\beta}}\qquad\text{and}\qquad \hat{A}_{\alpha\beta\gamma}=A_0\,\hat{Y}_{\alpha\beta\gamma}+\Delta \hat{A}_{\alpha\beta\gamma} \,,
\end{equation}
a solution to the supersymmetric flavour problem for low-energy sparticle spectra requires:
\begin{equation}
\Delta m^2_{\alpha\overline{\beta}} \ll m_0^2\,\delta_{\alpha\overline{\beta}}    \qquad\text{and}\qquad 
A_0\,\hat{Y}_{\alpha\beta\gamma} \ll \hat{A}_{\alpha\beta\gamma}\,.
\label{FlavourCond}
\end{equation}
These conditions translate into strong bounds once observations are taken into account. A very constraining process is $K^0$-$\overline{K}^0$ mixing which receives new contributions from squark loops, leading to the following bound \cite{Luty:2005sn}:
\begin{equation}
\frac{\Delta m_{\tilde{s}\tilde{d}}^2}{m_0^2}\lesssim V_{ts}V_{td}\left(\frac{m_0}{M_W}\right)\sim 10^{-3}\left(\frac{m_0}{500 \text{GeV}}\right),
\label{eq:FCNCc}
\end{equation}
which shows that TeV-scale squarks require a hierarchy of at least three orders of magnitude between off-diagonal and diagonal entries of the scalar mass-squared matrix. Note that no relevant bound arises if instead scalar masses are pushed around $10^6$ GeV or higher.

The condition \eqref{eq:FCNCc} can be extended to all FCNC contributions induced by the soft-terms. Focusing on the squark sector, it is useful to define the dimensionless quantity \cite{Isidori:2010kg}:
\begin{equation}
(\delta_{\alpha\overline{\beta}}^{\tilde{q}})_{MN}\equiv \frac{(M_{\tilde{q}}^2)^{MN}_{\alpha\overline{\beta}}}{m_0^2}\,,
\label{eq:FCNCc2}
\end{equation}
where $M,N=L,R$ label the chirality, $M_{\tilde{q}}^2$ denotes the squark mass-squared matrix, and $m_0^2$ is the average squark mass-squared. More precisely, using our notation, $(M_{\tilde{q}}^2)^{MM}_{\alpha\overline{\beta}} \equiv (m_{\tilde{q}}^2)^{MM}_{\alpha\overline{\beta}}$ and $(M_{\tilde{q}}^2)^{LR}_{\alpha\overline{\beta}} \equiv \hat{A}_{q\alpha\beta}\langle H^0_q\rangle$. The absence of dangerously large FCNCs for TeV-scale supersymmetry corresponds to requiring $(\delta_{\alpha\overline{\beta}}^{\tilde{q}})_{MM}\ll 1$ with $\alpha\neq \beta$, i.e. small off-diagonal terms of the soft scalar masses, and $(\delta_{\alpha\overline{\beta}}^{\tilde{q}})_{LR}\ll 1$, which is clearly satisfied if the trilinear A-terms are proportional to the physical Yukawa couplings. The tightest of these bounds comes from the Electric Dipole Moment (EDM) of the neutron (using the relation between the neutron and quark EDMs) and reads \cite{Gabbiani:1996hi}: 
\begin{equation}
(\delta_{1\overline{1}}^d)_{LR}< 4.7\times 10^{-6}\left(\frac{m_0}{1\,\text{TeV}} \right)^2\left(\frac{1\,\text{TeV}}{M_{1/2}} \right).
\label{eq:FCNCc3}
\end{equation}

Similarly to the quark sector, flavour violation in the lepton sector is also constrained by several experiments. Extending the definition (\ref{eq:FCNCc2}) to the slepton sector, the most severe bound comes from the electron EDM which gives:\footnote{Let us stress that this bound applies only if the axionic partners of the moduli develop appropriate vacuum expectation values so that the imaginary part of the A-terms is non-zero.}
\begin{equation}
|\text{Im}(\delta_{1\overline{1}}^l)_{LR}|< 5.8\times 10^{-9}\left(\frac{m_0}{1\,\text{TeV}} \right)^2\left(\frac{1\,\text{TeV}}{M_{1/2}} \right),
\label{eq:FCNCc4}
\end{equation}
where we used the relevant formula in \cite{Gabbiani:1996hi} and the current experimental limit by ACME collaboration at 90\% confidence level \cite{ACME:2018yjb}:
\begin{equation}
    \biggl|\frac{d_e}{e}\biggl| \lesssim 1.1 \times 10^{-29}\,\text{cm} = 5.6\times 10^{-13}\,\text{TeV}^{-1}\,.
\end{equation}
This upper bound is expected to be pushed by ACME III to $1.5\times 10^{-14}\,\text{TeV}^{-1}$ \cite{Kara:2012ay,ACMEIII}.\footnote{Note that such experimental limits constrain also possible higher-dimensional operators in the SM effective field theory with flavour symmetry, as recently discussed in \cite{Kobayashi:2021pav,Kobayashi:2022jvy}.}

It is worth mentioning that the trilinear A-terms are constrained, not just by the flavour supersymmetric problem, but also from the requirement to avoid charge and colour breaking vacua and an unbounded from below potential \cite{Casas:1996de,Casas:1995pd}. The associated bounds are in general stricter than the ones from FCNCs and do not depend on the scale of supersymmetry breaking. They take the form:
\begin{eqnarray}
\left|\hat{A}_{u\alpha\beta}\right|^2&\leq&
\hat{Y}_{u\gamma\gamma}^2\left(m_{u_{L\alpha}}^2+m_{u_{R\beta}}^2+m_{H_u}^2\right), \quad\gamma=\text{max}(\alpha,\beta)\,, \nonumber \\
\left|\hat{A}_{d\alpha\beta}\right|^2 &\leq& \hat{Y}_{d\gamma\gamma}^2\left(m_{d_{L\alpha}}^2+m_{d_{R\beta}}^2+m_{H_d}^2\right), \quad\gamma=\text{max}(\alpha,\beta)\,,
\label{eq:UFBc} \\
\left|\hat{A}_{l\alpha\beta}\right|^2&\leq& \hat{Y}_{e\gamma\gamma}^2\left(m_{e_{L\alpha}}^2+m_{e_{R\beta}}^2+m_{H_d}^2\right), \quad\gamma=\text{max}(\alpha,\beta)\,. \nonumber
\end{eqnarray}

\subsection{Mirror mediation}
\label{subsec:Mirror Mediation}

In gravity mediation supersymmetry breaking, the experimental bounds described in Sec. \ref{ExpConstr} are in general not automatically satisfied. The conditions to break supersymmetry without generating large FCNCs are called \emph{mirror mediation} and have been derived in \cite{Conlon:2007dw} finding:
\begin{enumerate}
\item The hidden sector fields have to be factorised into two classes which we will collectively denote as $\Psi$ and $\chi$.

\item There should be no kinetic mixing between $\Phi$ and $\chi$, i.e. the K\"ahler potential should be a direct sum of two terms where the first depends just on the real part of $\Psi$:
\begin{equation}
K(\Psi+\overline{\Psi},\chi,\overline{\chi})=K_1(\Psi+\overline{\Psi})+K_2(\chi,\overline{\chi}).
\end{equation}
The reality condition on $\Psi$ is necessary to avoid relative phases between different A-terms.

\item The superpotential Yukawa couplings should depend just on the $\chi$ fields, while the gauge kinetic functions should depend (linearly) just on the $\Psi$ fields:
\begin{equation}
Y_{\alpha\beta\gamma}(\Psi,\chi)=Y_{\alpha\beta\gamma}(\chi)
\qquad\text{and}\qquad \mathfrak{f}_{a}(\Psi,\chi)=\sum_i\lambda_i \Psi_i\,.
\end{equation}
The linearity of $\mathfrak{f}_a$ allows to have universal gaugino mass phases which are aligned with those of the A-terms.

\item The K\"ahler matter metric should factorise as follows:
\begin{equation}
\tilde{K}_{\alpha\overline{\beta}}(\Psi,\overline{\Psi},\chi,\overline{\chi})=h(\Psi+\overline{\Psi})\,k_{\alpha\overline{\beta}}(\chi,\overline{\chi})\,.
\end{equation}

\item Supersymmetry breaking should be induced by the $\Psi$ fields, while the $\chi$ fields should be stabilised supersymmetrically:
\begin{equation}
F^\Psi\neq 0 \qquad\text{and}\qquad F^\chi=0\,,
\end{equation}
which, given assumption 2, is also equivalent to:
\begin{equation}
D_{\Psi}W\neq 0 \qquad\text{and}\qquad D_{\chi}W=0\,.
\end{equation}
\end{enumerate}
These conditions generate flavour universal soft-terms since they imply the existence of two decoupled sectors: the $\Phi$ fields which are responsible to break supersymmetry, and the $\chi$ fields which generate the flavour structure. One can see that these conditions indeed induce the flavour universal conditions (\ref{FlavourCond}) by looking at the unnormalised scalar masses and trilinear A-terms:
\begin{eqnarray}
\tilde{m}_{\alpha\overline{\beta}}^2&=& m_{3/2}^2 K_{\alpha\overline{\beta}}-\overline{F}^{\overline{\Psi}}F^{\Psi}\left(\partial_{\overline{\Psi}}\partial_{\Psi}K_{\alpha\overline{\beta}}-(\partial_{\overline{\Psi}}K_{\alpha\overline{\gamma}})K^{\overline{\gamma}\delta}(\partial_{\Psi}K_{\delta\overline{\beta}})\right)
\nonumber\\
&=& \left(m_{3/2}^2 h-\overline{F}^{\overline{\Psi}}F^{\Psi}\left(\partial_{\overline{\Psi}}\partial_{\Psi}h-\frac{\partial_{\overline{\Psi}}h\partial_{\Psi}h}{h}\right)\right)k_{\alpha\overline{\beta}}(\chi,\overline{\chi})\,,
\nonumber\\
\tilde{A}_{\alpha\beta\gamma}&=&e^{K/2}\,Y_{\alpha\beta\gamma}(\chi)\left(F^{\Psi}\partial_{\Psi}K(\Psi,\overline{\Psi})-3\frac{F^{\Psi}\partial_{\Psi}h(\Psi,\overline{\Psi})}{h(\Psi,\overline{\Psi})}\right).
\end{eqnarray}
These seem to be \emph{ad hoc} conditions to obtain flavour universal soft terms. However, as already pointed out in \cite{Conlon:2007dw}, they are satisfied in the large class of type IIB string vacua where the MSSM lives on D7-branes. In the next section we will study if mirror mediation can occur also in type IIB sequestered models with 
the MSSM on D3-branes.

\subsection{Sequestering and mirror mediation}
\label{subsec:Sequestering and Mirror Mediation}

In this section, we discuss if the soft-terms obtained in Sec. \ref{sec:Sequestered Supersymmetry Breaking} respect the five mirror mediation conditions described above:
\begin{enumerate}
\item The hidden sector factorises into two classes. The first includes the dilaton $S$ and the complex structure moduli $U$, whereas the second involves the K\"ahler moduli $T_b$, $T_s$, $T_{\mathrm{dS}}$, $T_{\mathrm{SM}}$ and $G$:
\begin{itemize}
\item $\chi=\{S,\,U\}$.
\item $\Phi=\{T_b,\,T_s,\,T_{\mathrm{dS}},\,T_{\mathrm{SM}},\,G\}$;
\end{itemize}
Hence the model matches the first condition.

\item The K\"ahler potential factorises at leading order but $\alpha'$ corrections induce a mixing between the $\Phi$ and $\chi$ sectors:
\begin{equation}
K=\underbrace{K_{cs}(U,\bar{U})-\ln(S+\overline{S})}_\text{$\chi$ sector}\underbrace{-2\ln \mathcal{V}+\lambda_{\mathrm{dS}}\frac{\tau_{\mathrm{dS}}^2}{\mathcal{V}}+\lambda_{\mathrm{SM}}\frac{\tau_{\mathrm{SM}}^2}{\mathcal{V}}+\lambda_b\frac{b^2}{\mathcal{V}}}_{\text{$\Phi$ sector}}\underbrace{-\frac{\xi}{\mathcal{V}}\left(\frac{S+\overline{S}}{2}\right)^{3/2}}_\text{mixing term}.
\end{equation}
Thus the second condition is respected only at leading order in the large volume expansion.

\item The mirror mediation condition on the superpotential Yukawa couplings is respected since they depend only on the $\chi$ fields due to the shift symmetry of the K\"ahler moduli and the holomorphy of the superpotential. On the other hand, the gauge kinetic functions do not respect the third condition as they depend on both $\Phi$ and $\chi$:
\begin{equation}
\mathfrak{f}_a=\delta_aS+\kappa_aT_{\mathrm{SM}}\,.
\end{equation}

\item Similarly to condition 2, the matter K\"ahler potential factorises just at leading order since $\alpha'$ corrections induce a mixing between the $\Phi$ and $\chi$ sectors (setting $\tau_{\mathrm{SM}}=\tau_{\mathrm{dS}}=b=0$):
\begin{equation}
\tilde{K}_{\alpha\overline{\beta}}=\underbrace{f_{\alpha\overline{\beta}}(U,\overline{U},S,\overline{S})}_\text{$\chi$ factor}\underbrace{\frac{1}{\mathcal{V}^{2/3}}}_\text{$\Phi$ factor}\left(1-\underbrace{c_s\frac{\xi}{\mathcal{V}}\left(\frac{S+\overline{S}}{2}\right)^{3/2}}_\text{mixing term}\right).
\end{equation}
Hence the fourth mirror mediation assumption is not respected by higher order corrections.

\item The fifth condition is badly violated in sequestered models since the F-terms of the two local $\Phi$ fields, $T_{\mathrm{SM}}$ and $G$, vanish: $F^{T_{\mathrm{SM}}}=F^G=0$. The F-terms of the other K\"ahler moduli are instead non-zero and scale as $F^{T_b}/\tau_b\sim F^{T_s}\sim F^{T_{\mathrm{dS}}}\sim \mathcal{O}\left(\mathcal{V}^{-1}\right)$. Moreover, the $\chi$ fields develop non-zero F-terms since the second condition is not perfectly matched. However, given that the kinetic mixing is induced only at subleading order, the associated F-terms are volume suppressed: $F^S\sim F^U\sim \mathcal{O}\left(\mathcal{V}^{-2}\right)$.
\end{enumerate}
This analysis shows that the mirror mediation conditions are in general satisfied at leading order, apart from the gauge kinetic functions and the vanishing of the F-terms of $T_{\mathrm{SM}}$ and $G$. One could therefore be tempted to conclude that sequestered D3-brane models might give rise to universal soft-terms at leading order. As we have seen in Sec. \ref{sec:Sequestered Supersymmetry Breaking} this is however in general not true. The reason is that the locality of the D3-brane construction induces a cancellation of the contribution to the soft-terms of the F-terms of the K\"ahler moduli. Combined with the fact that $F^{T_{\mathrm{SM}}}=F^G=0$, the main contribution to the scalar masses and the trilinear A-terms can then come from the F-terms of $S$ and $U$ which control also the superpotential Yukawa couplings. This is indeed the case in the ultra-local limit which is therefore expected to be severely constrained by FCNCs. On the other hand, in the local limit the cancellation of the contribution to the soft-terms of $F^{T_b}$ is not perfect, allowing to match the mirror mediation conditions at leading order. This is not true for the A-terms but the hierarchy between them and the scalar masses can easily suppress large FCNCs.

\subsection{Comparison with experiments}

After the qualitative discussion of the previous section, let us now check in a more quantitative way if the soft-terms computed in Sec. \ref{sec:Sequestered Supersymmetry Breaking} respect the experimental bounds described in Sec. \ref{FlavourCond}. We shall consider the local and ultra-local limits separately.

\subsubsection{Local limit}
\label{subsubsec:Local Scenario}

As we have seen in Sec. \ref{sec:Sequestered Supersymmetry Breaking}, in the local scenario scalar masses are nearly flavour universal with the leading diagonal result given by (\ref{m0Local}) and subdominant off-diagonal contributions given by (\ref{m0UltraLoc}):
\begin{equation}
m_0^2=\left(c_s-\frac{1}{3}\right)\frac{5}{\omega_S'}\,m_{3/2}M_{1/2}\qquad\text{and}\qquad
\Delta m_{\alpha\overline{\beta}}^2=g_{\alpha\overline{\beta}}\, M_{1/2}^2\,,
\end{equation}
where $g_{\alpha\overline{\beta}}(S,\overline{S},U,\overline{U})$ is an $\mathcal{O}(1)$ function of the dilaton and the complex structure moduli. The model is therefore safe from large FCNCs. In fact, applying the experimental bound (\ref{eq:FCNCc}) we find:\footnote{Note that our bounds differ from the ones obtained in \cite{deAlwis:2009fn,deAlwis:2012vp} which assumed that the K\"ahler metric for matter fields on D3-branes at singularities takes the same form as in the case where the D3-branes are at a smooth point.}
\begin{equation}
\mathcal{V}\gtrsim 10^{-2} \left[\frac{\hat{\xi}\,\omega_S'^2\,g_{\alpha\overline{\beta}}}{\left(c_s-\frac{1}{3}\right)}\right]\left( \frac{10^7\,\text{GeV}}{m_0}\right),
\label{ImpBound}
\end{equation}
which is clearly satisfied for $\mathcal{V}\sim 10^7$ that yields $m_0\sim 10^7$ GeV since the factor in square brackets is expected to be of order unity. Expressing generically $m_0$ in terms of the volume using (\ref{m0Local}), the bound (\ref{ImpBound}) reduces to $\mathcal{V}\lesssim 10^{26}$ which is always satisfied for $\mathcal{V} \lesssim 10^7$ that forbids gaugino masses below the TeV-scale. It is straightforward to check that also the constraints (\ref{eq:FCNCc3}) and (\ref{eq:FCNCc4}) are easily satisfied in the local scenario thanks to the hierarchy between scalar and gaugino masses. Moreover, the CCB and UFB conditions (\ref{eq:UFBc}) are automatically satisfied in the local scenario since $\hat{A}_{\alpha\beta\gamma}\sim M_{1/2}\ll m_0$.

Let us finally comment on the fact that the split supersymmetry spectrum of the local scenario affects the value of the mass of the Standard Model Higgs. Ref. \cite{Bagnaschi:2014rsa} computed the renormalisation group flow of the Higgs mass in the MSSM for different supersymmetry breaking scenarios. For split supersymmetry with $M_{1/2}\sim 1$ TeV and $m_0\sim 10^7$ GeV, matching the observed value of the Higgs mass requires $\tan\beta\simeq 1$. 

\subsubsection{Ultra-local limit}
\label{subsubsec:Ultra-local Scenario}

In the ultra-local case squark and slepton masses are given by (\ref{m0UltraLoc}):
\begin{equation}
m_{\alpha\overline{\beta}}^2=g_{\alpha\overline{\beta}}(S,\overline{S},U,\overline{U})\, M_{1/2}^2\,.
\end{equation}
This expression is not flavour universal and there is no clear source of suppression of the off-diagonal terms. Given that $g_{\alpha\overline{\beta}}$ is expected to be an $\mathcal{O}(1)$ function of the dilaton and the complex structure moduli, the LHS of the bound (\ref{eq:FCNCc}) is also expected to be of $\mathcal{O}(1)$. Moreover, in the ultra-local case scalar masses scale as gaugino masses, and so TeV-scale soft-terms would violate the bound (\ref{eq:FCNCc}) since the term on the RHS would be of $\mathcal{O}(10^{-3})$. The only way to satisfy the bound (\ref{eq:FCNCc}) seems therefore to push all the soft-terms to higher scales: $M_{\mathrm{soft}}\gtrsim 10^6$ GeV. 

A similar lower bound on the soft-terms comes from the condition (\ref{eq:FCNCc4}) which involves the trilinear A-terms that in the ultra-local case are in general not proportional to the physical Yukawa couplings. Thus they are more generically proportional to a combination of the Yukawa couplings which is expected to scale as the largest Yukawa coupling of the $u$-, $d$- or $l$-type respectively, unless some hidden structure gives rise to a further suppression. Given that we do not see any reason for this to happen, we expect the relevant physical A-term to scale as $\hat{A}_{l11}\sim y_\tau\,M_{1/2}$, where $y_\tau\equiv \hat{Y}_{l33}$ denotes the $\tau$ Yukawa coupling. Assuming $|{\mathrm{Im}}(\delta^l_{1\overline{1}})_{LR}|\sim |(\delta^l_{1\overline{1}})_{LR}|$, and taking $m_0\sim M_{1/2}$, the bound (\ref{eq:FCNCc4}) becomes again $M_{\mathrm{soft}}\gtrsim 10^6$ GeV. Therefore the ultra-local limit can be compatible with flavour observables if supersymmetry shows up at high scales. 

However the CCB/UFB conditions (\ref{eq:UFBc}) are much more stringent since they cannot be evaded by simply enhancing the mass scale of the soft-terms. Given that squark and slepton masses scale as the A-terms which are not proportional to the physical Yukawa couplings, only the CCB/UFB conditions involving the heaviest sleptons and $u$- and $d$-type squarks seem to be satisfiable. On the other hand, the most dangerous conditions are the ones for the trilinear A-terms associated to the lightest sfermions since we expect them to scale as the Yukawa couplings of the heaviest fermions. For example, as already explained, we expect $\hat{A}_{l11}\sim y_\tau\,M_{1/2}\sim y_\tau\,m_0$  which would not satisfy the third condition in (\ref{eq:UFBc}) since $y_e \ll y_\tau$ (where $y_e$ is the electron Yukawa coupling). We conclude that, unless unexpected new cancellations and suppressions arise from the complex structure sector, the ultra-local scenario does not seem to survive CCB/UFB constraints, leading to unstable MSSM models.

Let us however point out that the inclusion of Planck-suppressed operators from UV physics, or radiative corrections to the scalar potential of the form $\Delta V \simeq \mathrm{Str} M^4 \ln\left(\frac{M^2}{\Lambda^2}\right)$, could in principle yield new stable minima without UFB directions. Studying these effects in detail is however beyond the scope of our paper. Moreover, we could very well coexist with CCB minima, as long as they are not populated in the early Universe and the lifetime of the metastable Standard Model-like vacuum is longer than the age of the Universe to prevent its decay to CCB minima. Interestingly, as shown in \cite{Kusenko:1996jn}, this is indeed the case for high scale supersymmetry, which we are requiring anyway to avoid large FCNCs. We will not try to envisage a cosmological mechanism which could guarantee the fact that CCB minima are not populated in the early Universe, but we will just note that this does not seem to be the generic situation. We thus conclude that sequestered models with an MSSM-like spectrum are in severe tension with stability bounds.

\section{Conclusions}
\label{Concl}

The low-energy limit of type IIB string compactifications on Calabi-Yau orientifolds is an $N=1$ supergravity theory. Background fluxes combined with quantum corrections beyond the tree-level approximation can stabilise all closed string moduli whose F-terms break supersymmetry spontaneously. This breaking is mediated to the MSSM living on D-branes via gravitational interactions, resulting in the generation of non-zero soft-terms. When the MSSM is realised with D7-branes, the soft-terms scale as the gravitino mass, $M_{\mathrm{soft}}\sim m_{3/2}$, while D3-branes can lead to sequestered models with $M_{\mathrm{soft}}\ll m_{3/2}$. 

This hierarchy can be very useful for phenomenological applications since it can allow for 4D string models with a large Hubble scale during inflation, low-energy supersymmetry, and no cosmological moduli problem. In this paper we therefore focused on LVS sequestered string models and analysed under what conditions they can also satisfy current flavour bounds. When gaugino and sfermion masses are of the same order of magnitude, FCNC constraints can be satisfied by raising the energy scale of the soft-terms to $M_{\mathrm{soft}}\gtrsim 10^6$ GeV. However, regardless of the value of $M_{\mathrm{soft}}$, these models are in tension with data due to the emergence of charge and colour breaking vacua together with potential directions along which the potential is unbounded from below.

On the other hand, sequestering is compatible with split supersymmetry scenarios where gauginos are hierarchically lighter than squarks and sleptons. In this case, supersymmetry breaking does not yield large FCNCs since scalar masses are flavour diagonal at leading order and the trilinear A-terms, even if not proportional to the physical Yukawa couplings, are suppressed with respect to scalar masses. This last feature guarantees also the absence of CCB vacua and UFB directions. Phenomenologically viable sequestered LVS models are therefore characterised by a split supersymmetry spectrum. For example, $\mathcal{V}\sim 10^7$ would give rise to TeV-scale gauginos, a lightest modulus (the volume mode) and sfermions around $10^7$ GeV, which requires $\tan\beta\simeq 1$ to reproduce the correct Higgs mass, and a gravitino mass around $10^{10}$ GeV. 

Note that this scenario can arise in different cases: ($i$) in the \emph{local} limit of the K\"ahler metric for matter fields for any uplifting mechanism except when this is achieved by non-zero F-terms of the complex structure moduli \cite{Gallego:2017dvd} since this case can lead to CCB/UFB vacua given that $M_{1/2}\sim m_0$; ($ii$) in the \emph{ultra-local} limit any time the uplifting sector induces large D-term contributions to scalar masses that lead to $M_{1/2}\ll m_0$, as for T-brane uplifting \cite{Cicoli:2015ylx}. Let us stress that this result is also supported by model building considerations. In fact, it appears generic from a top-down perspective since the ultra-local limit seems hard to realise given that it  would require a cancellation of the volume dependence of the physical Yukawa couplings at all orders. Moreover, even in the ultra-local limit, the uplifting sector should also be localised to suppress its D-term contribution to scalar masses, as for dilaton-dependent non-perturbative effects at singularities \cite{Cicoli:2012fh}, which again does not seem a generic situation. Let us finally mention that this analysis leads to a clear phenomenological prediction since it indicates that collider searches for supersymmetric particles should discover gauginos before squarks and sleptons.

\acknowledgments

This work was supported by JSPS KAKENHI Grant Numbers JP20K14477 (H. O.) and JP23H04512 (H. O.). AC's research is funded by the Deutsche Forschungsgemeinschaft (DFG, German Research Foundation) - Projektnummer 417533893/GRK2575 “Rethinking Quantum Field Theory”.

\bibliographystyle{JHEP} 
\bibliography{draft}

\providecommand{\href}[2]{#2}\begingroup\raggedright\begin{thebibliography}{10}

\bibitem{Balasubramanian:2005zx}
V.~Balasubramanian, P.~Berglund, J.P.~Conlon and F.~Quevedo, \emph{{Systematics
  of moduli stabilisation in Calabi-Yau flux compactifications}},
  \href{https://doi.org/10.1088/1126-6708/2005/03/007}{\emph{JHEP} {\bfseries
  03} (2005) 007} [\href{https://arxiv.org/abs/hep-th/0502058}{{\ttfamily
  hep-th/0502058}}].

\bibitem{Cicoli:2008va}
M.~Cicoli, J.P.~Conlon and F.~Quevedo, \emph{{General Analysis of LARGE Volume
  Scenarios with String Loop Moduli Stabilisation}},
  \href{https://doi.org/10.1088/1126-6708/2008/10/105}{\emph{JHEP} {\bfseries
  10} (2008) 105} [\href{https://arxiv.org/abs/0805.1029}{{\ttfamily
  0805.1029}}].

\bibitem{Conlon:2005ki}
J.P.~Conlon, F.~Quevedo and K.~Suruliz, \emph{{Large-volume flux
  compactifications: Moduli spectrum and D3/D7 soft supersymmetry breaking}},
  \href{https://doi.org/10.1088/1126-6708/2005/08/007}{\emph{JHEP} {\bfseries
  08} (2005) 007} [\href{https://arxiv.org/abs/hep-th/0505076}{{\ttfamily
  hep-th/0505076}}].

\bibitem{Conlon:2006wz}
J.P.~Conlon, S.S.~Abdussalam, F.~Quevedo and K.~Suruliz, \emph{{Soft SUSY
  Breaking Terms for Chiral Matter in IIB String Compactifications}},
  \href{https://doi.org/10.1088/1126-6708/2007/01/032}{\emph{JHEP} {\bfseries
  01} (2007) 032} [\href{https://arxiv.org/abs/hep-th/0610129}{{\ttfamily
  hep-th/0610129}}].

\bibitem{Blumenhagen:2009gk}
R.~Blumenhagen, J.P.~Conlon, S.~Krippendorf, S.~Moster and F.~Quevedo,
  \emph{{SUSY Breaking in Local String/F-Theory Models}},
  \href{https://doi.org/10.1088/1126-6708/2009/09/007}{\emph{JHEP} {\bfseries
  09} (2009) 007} [\href{https://arxiv.org/abs/0906.3297}{{\ttfamily
  0906.3297}}].

\bibitem{Aparicio:2014wxa}
L.~Aparicio, M.~Cicoli, S.~Krippendorf, A.~Maharana, F.~Muia and F.~Quevedo,
  \emph{{Sequestered de Sitter String Scenarios: Soft-terms}},
  \href{https://doi.org/10.1007/JHEP11(2014)071}{\emph{JHEP} {\bfseries 11}
  (2014) 071} [\href{https://arxiv.org/abs/1409.1931}{{\ttfamily 1409.1931}}].

\bibitem{Reece:2015qbf}
M.~Reece and W.~Xue, \emph{{SUSY\textquoteright{}s Ladder: reframing
  sequestering at Large Volume}},
  \href{https://doi.org/10.1007/JHEP04(2016)045}{\emph{JHEP} {\bfseries 04}
  (2016) 045} [\href{https://arxiv.org/abs/1512.04941}{{\ttfamily
  1512.04941}}].

\bibitem{Coughlan:1983ci}
G.D.~Coughlan, W.~Fischler, E.W.~Kolb, S.~Raby and G.G.~Ross,
  \emph{{Cosmological Problems for the Polonyi Potential}},
  \href{https://doi.org/10.1016/0370-2693(83)91091-2}{\emph{Phys. Lett. B}
  {\bfseries 131} (1983) 59}.

\bibitem{Banks:1993en}
T.~Banks, D.B.~Kaplan and A.E.~Nelson, \emph{{Cosmological implications of
  dynamical supersymmetry breaking}},
  \href{https://doi.org/10.1103/PhysRevD.49.779}{\emph{Phys. Rev. D} {\bfseries
  49} (1994) 779} [\href{https://arxiv.org/abs/hep-ph/9308292}{{\ttfamily
  hep-ph/9308292}}].

\bibitem{deCarlos:1993wie}
B.~de~Carlos, J.A.~Casas, F.~Quevedo and E.~Roulet, \emph{{Model independent
  properties and cosmological implications of the dilaton and moduli sectors of
  4-d strings}},
  \href{https://doi.org/10.1016/0370-2693(93)91538-X}{\emph{Phys. Lett. B}
  {\bfseries 318} (1993) 447}
  [\href{https://arxiv.org/abs/hep-ph/9308325}{{\ttfamily hep-ph/9308325}}].

\bibitem{Bertolini:1990if}
S.~Bertolini, F.~Borzumati, A.~Masiero and G.~Ridolfi, \emph{{Effects of
  supergravity induced electroweak breaking on rare $B$ decays and mixings}},
  \href{https://doi.org/10.1016/0550-3213(91)90320-W}{\emph{Nucl. Phys. B}
  {\bfseries 353} (1991) 591}.

\bibitem{Gabbiani:1996hi}
F.~Gabbiani, E.~Gabrielli, A.~Masiero and L.~Silvestrini, \emph{{A Complete
  analysis of FCNC and CP constraints in general SUSY extensions of the
  standard model}},
  \href{https://doi.org/10.1016/0550-3213(96)00390-2}{\emph{Nucl. Phys. B}
  {\bfseries 477} (1996) 321}
  [\href{https://arxiv.org/abs/hep-ph/9604387}{{\ttfamily hep-ph/9604387}}].

\bibitem{Isidori:2010kg}
G.~Isidori, Y.~Nir and G.~Perez, \emph{{Flavor Physics Constraints for Physics
  Beyond the Standard Model}},
  \href{https://doi.org/10.1146/annurev.nucl.012809.104534}{\emph{Ann. Rev.
  Nucl. Part. Sci.} {\bfseries 60} (2010) 355}
  [\href{https://arxiv.org/abs/1002.0900}{{\ttfamily 1002.0900}}].

\bibitem{Misiak:1997ei}
M.~Misiak, S.~Pokorski and J.~Rosiek, \emph{{Supersymmetry and FCNC effects}},
  \href{https://doi.org/10.1142/9789812812667_0012}{\emph{Adv. Ser. Direct.
  High Energy Phys.} {\bfseries 15} (1998) 795}
  [\href{https://arxiv.org/abs/hep-ph/9703442}{{\ttfamily hep-ph/9703442}}].

\bibitem{Abe:2011rg}
H.~Abe, H.~Otsuka, Y.~Sakamura and Y.~Yamada, \emph{{SUSY Flavor Structure of
  Generic 5D Supergravity Models}},
  \href{https://doi.org/10.1140/epjc/s10052-012-2018-x}{\emph{Eur. Phys. J. C}
  {\bfseries 72} (2012) 2018}
  [\href{https://arxiv.org/abs/1111.3721}{{\ttfamily 1111.3721}}].

\bibitem{Casas:1996de}
J.A.~Casas and S.~Dimopoulos, \emph{{Stability bounds on flavor violating
  trilinear soft terms in the MSSM}},
  \href{https://doi.org/10.1016/0370-2693(96)01000-3}{\emph{Phys. Lett. B}
  {\bfseries 387} (1996) 107}
  [\href{https://arxiv.org/abs/hep-ph/9606237}{{\ttfamily hep-ph/9606237}}].

\bibitem{Casas:1995pd}
J.A.~Casas, A.~Lleyda and C.~Munoz, \emph{{Strong constraints on the parameter
  space of the MSSM from charge and color breaking minima}},
  \href{https://doi.org/10.1016/0550-3213(96)00194-0}{\emph{Nucl. Phys. B}
  {\bfseries 471} (1996) 3}
  [\href{https://arxiv.org/abs/hep-ph/9507294}{{\ttfamily hep-ph/9507294}}].

\bibitem{Conlon:2007dw}
J.P.~Conlon, \emph{{Mirror Mediation}},
  \href{https://doi.org/10.1088/1126-6708/2008/03/025}{\emph{JHEP} {\bfseries
  03} (2008) 025} [\href{https://arxiv.org/abs/0710.0873}{{\ttfamily
  0710.0873}}].

\bibitem{Bagnaschi:2014rsa}
E.~Bagnaschi, G.F.~Giudice, P.~Slavich and A.~Strumia, \emph{{Higgs Mass and
  Unnatural Supersymmetry}},
  \href{https://doi.org/10.1007/JHEP09(2014)092}{\emph{JHEP} {\bfseries 09}
  (2014) 092} [\href{https://arxiv.org/abs/1407.4081}{{\ttfamily 1407.4081}}].

\bibitem{Kusenko:1996jn}
A.~Kusenko, P.~Langacker and G.~Segre, \emph{{Phase transitions and vacuum
  tunneling into charge and color breaking minima in the MSSM}},
  \href{https://doi.org/10.1103/PhysRevD.54.5824}{\emph{Phys. Rev. D}
  {\bfseries 54} (1996) 5824}
  [\href{https://arxiv.org/abs/hep-ph/9602414}{{\ttfamily hep-ph/9602414}}].

\bibitem{Cicoli:2012vw}
M.~Cicoli, S.~Krippendorf, C.~Mayrhofer, F.~Quevedo and R.~Valandro,
  \emph{{D-Branes at del Pezzo Singularities: Global Embedding and Moduli
  Stabilisation}}, \href{https://doi.org/10.1007/JHEP09(2012)019}{\emph{JHEP}
  {\bfseries 09} (2012) 019} [\href{https://arxiv.org/abs/1206.5237}{{\ttfamily
  1206.5237}}].

\bibitem{Cicoli:2013mpa}
M.~Cicoli, S.~Krippendorf, C.~Mayrhofer, F.~Quevedo and R.~Valandro,
  \emph{{D3/D7 Branes at Singularities: Constraints from Global Embedding and
  Moduli Stabilisation}},
  \href{https://doi.org/10.1007/JHEP07(2013)150}{\emph{JHEP} {\bfseries 07}
  (2013) 150} [\href{https://arxiv.org/abs/1304.0022}{{\ttfamily 1304.0022}}].

\bibitem{Cicoli:2013cha}
M.~Cicoli, D.~Klevers, S.~Krippendorf, C.~Mayrhofer, F.~Quevedo and
  R.~Valandro, \emph{{Explicit de Sitter Flux Vacua for Global String Models
  with Chiral Matter}},
  \href{https://doi.org/10.1007/JHEP05(2014)001}{\emph{JHEP} {\bfseries 05}
  (2014) 001} [\href{https://arxiv.org/abs/1312.0014}{{\ttfamily 1312.0014}}].

\bibitem{Cicoli:2017shd}
M.~Cicoli, I.n.~Garc\`\i{}a-Etxebarria, C.~Mayrhofer, F.~Quevedo, P.~Shukla and
  R.~Valandro, \emph{{Global Orientifolded Quivers with Inflation}},
  \href{https://doi.org/10.1007/JHEP11(2017)134}{\emph{JHEP} {\bfseries 11}
  (2017) 134} [\href{https://arxiv.org/abs/1706.06128}{{\ttfamily
  1706.06128}}].

\bibitem{Cicoli:2021dhg}
M.~Cicoli, I.n.G.~Etxebarria, F.~Quevedo, A.~Schachner, P.~Shukla and
  R.~Valandro, \emph{{The Standard Model quiver in de Sitter string
  compactifications}},
  \href{https://doi.org/10.1007/JHEP08(2021)109}{\emph{JHEP} {\bfseries 08}
  (2021) 109} [\href{https://arxiv.org/abs/2106.11964}{{\ttfamily
  2106.11964}}].

\bibitem{Cicoli:2015ylx}
M.~Cicoli, F.~Quevedo and R.~Valandro, \emph{{De Sitter from T-branes}},
  \href{https://doi.org/10.1007/JHEP03(2016)141}{\emph{JHEP} {\bfseries 03}
  (2016) 141} [\href{https://arxiv.org/abs/1512.04558}{{\ttfamily
  1512.04558}}].

\bibitem{Gallego:2017dvd}
D.~Gallego, M.C.D.~Marsh, B.~Vercnocke and T.~Wrase, \emph{{A New Class of de
  Sitter Vacua in Type IIB Large Volume Compactifications}},
  \href{https://doi.org/10.1007/JHEP10(2017)193}{\emph{JHEP} {\bfseries 10}
  (2017) 193} [\href{https://arxiv.org/abs/1707.01095}{{\ttfamily
  1707.01095}}].

\bibitem{Cicoli:2012fh}
M.~Cicoli, A.~Maharana, F.~Quevedo and C.P.~Burgess, \emph{{De Sitter String
  Vacua from Dilaton-dependent Non-perturbative Effects}},
  \href{https://doi.org/10.1007/JHEP06(2012)011}{\emph{JHEP} {\bfseries 06}
  (2012) 011} [\href{https://arxiv.org/abs/1203.1750}{{\ttfamily 1203.1750}}].

\bibitem{Gukov:1999ya}
S.~Gukov, C.~Vafa and E.~Witten, \emph{{CFT's from Calabi-Yau four folds}},
  \href{https://doi.org/10.1016/S0550-3213(00)00373-4}{\emph{Nucl. Phys. B}
  {\bfseries 584} (2000) 69}
  [\href{https://arxiv.org/abs/hep-th/9906070}{{\ttfamily hep-th/9906070}}].

\bibitem{Grana:2003ek}
M.~Grana, T.W.~Grimm, H.~Jockers and J.~Louis, \emph{{Soft supersymmetry
  breaking in Calabi-Yau orientifolds with D-branes and fluxes}},
  \href{https://doi.org/10.1016/j.nuclphysb.2004.04.021}{\emph{Nucl. Phys. B}
  {\bfseries 690} (2004) 21}
  [\href{https://arxiv.org/abs/hep-th/0312232}{{\ttfamily hep-th/0312232}}].

\bibitem{Becker:2002nn}
K.~Becker, M.~Becker, M.~Haack and J.~Louis, \emph{{Supersymmetry breaking and
  alpha-prime corrections to flux induced potentials}},
  \href{https://doi.org/10.1088/1126-6708/2002/06/060}{\emph{JHEP} {\bfseries
  06} (2002) 060} [\href{https://arxiv.org/abs/hep-th/0204254}{{\ttfamily
  hep-th/0204254}}].

\bibitem{Lust:2004cx}
D.~Lust, P.~Mayr, R.~Richter and S.~Stieberger, \emph{{Scattering of gauge,
  matter, and moduli fields from intersecting branes}},
  \href{https://doi.org/10.1016/j.nuclphysb.2004.06.052}{\emph{Nucl. Phys. B}
  {\bfseries 696} (2004) 205}
  [\href{https://arxiv.org/abs/hep-th/0404134}{{\ttfamily hep-th/0404134}}].

\bibitem{Lust:2004fi}
D.~Lust, S.~Reffert and S.~Stieberger, \emph{{Flux-induced soft supersymmetry
  breaking in chiral type IIB orientifolds with D3 / D7-branes}},
  \href{https://doi.org/10.1016/j.nuclphysb.2004.11.030}{\emph{Nucl. Phys. B}
  {\bfseries 706} (2005) 3}
  [\href{https://arxiv.org/abs/hep-th/0406092}{{\ttfamily hep-th/0406092}}].

\bibitem{Conlon:2006tj}
J.P.~Conlon, D.~Cremades and F.~Quevedo, \emph{{Kahler potentials of chiral
  matter fields for Calabi-Yau string compactifications}},
  \href{https://doi.org/10.1088/1126-6708/2007/01/022}{\emph{JHEP} {\bfseries
  01} (2007) 022} [\href{https://arxiv.org/abs/hep-th/0609180}{{\ttfamily
  hep-th/0609180}}].

\bibitem{Dudas:2005vv}
E.~Dudas and S.K.~Vempati, \emph{{Large D-terms, hierarchical soft spectra and
  moduli stabilisation}},
  \href{https://doi.org/10.1016/j.nuclphysb.2005.08.034}{\emph{Nucl. Phys. B}
  {\bfseries 727} (2005) 139}
  [\href{https://arxiv.org/abs/hep-th/0506172}{{\ttfamily hep-th/0506172}}].

\bibitem{Luty:2005sn}
M.A.~Luty, \emph{{2004 TASI lectures on supersymmetry breaking}},  in
  \emph{{Theoretical Advanced Study Institute in Elementary Particle Physics}:
  {Physics in D $\geqq$ 4}}, pp.~495--582, 9, 2005
  [\href{https://arxiv.org/abs/hep-th/0509029}{{\ttfamily hep-th/0509029}}].

\bibitem{Dimopoulos:1995ju}
S.~Dimopoulos and D.W.~Sutter, \emph{{The Supersymmetric flavor problem}},
  \href{https://doi.org/10.1016/0550-3213(95)00421-N}{\emph{Nucl. Phys. B}
  {\bfseries 452} (1995) 496}
  [\href{https://arxiv.org/abs/hep-ph/9504415}{{\ttfamily hep-ph/9504415}}].

\bibitem{deCarlos:1995ah}
B.~de~Carlos, J.A.~Casas and J.M.~Moreno, \emph{{Constraints on supersymmetric
  theories from mu ---\ensuremath{>} e gamma}},
  \href{https://doi.org/10.1103/PhysRevD.53.6398}{\emph{Phys. Rev. D}
  {\bfseries 53} (1996) 6398}
  [\href{https://arxiv.org/abs/hep-ph/9507377}{{\ttfamily hep-ph/9507377}}].

\bibitem{ACME:2018yjb}
{\scshape ACME} collaboration, \emph{{Improved limit on the electric dipole
  moment of the electron}},
  \href{https://doi.org/10.1038/s41586-018-0599-8}{\emph{Nature} {\bfseries
  562} (2018) 355}.

\bibitem{Kara:2012ay}
D.M.~Kara, I.J.~Smallman, J.J.~Hudson, B.E.~Sauer, M.R.~Tarbutt and E.A.~Hinds,
  \emph{{Measurement of the electron's electric dipole moment using YbF
  molecules: methods and data analysis}},
  \href{https://doi.org/10.1088/1367-2630/14/10/103051}{\emph{New J. Phys.}
  {\bfseries 14} (2012) 103051}
  [\href{https://arxiv.org/abs/1208.4507}{{\ttfamily 1208.4507}}].

\bibitem{ACMEIII}
{\scshape ACME} collaboration, \emph{{Search for the Electric Dipole Moment of
  the Electron with Thorium Monoxide - The ACME Experiment.}}, {\emph{Talk at
  the KITP, September} (2016) }.

\bibitem{Kobayashi:2021pav}
T.~Kobayashi, H.~Otsuka, M.~Tanimoto and K.~Yamamoto, \emph{{Modular symmetry
  in the SMEFT}},
  \href{https://doi.org/10.1103/PhysRevD.105.055022}{\emph{Phys. Rev. D}
  {\bfseries 105} (2022) 055022}
  [\href{https://arxiv.org/abs/2112.00493}{{\ttfamily 2112.00493}}].

\bibitem{Kobayashi:2022jvy}
T.~Kobayashi, H.~Otsuka, M.~Tanimoto and K.~Yamamoto, \emph{{Lepton flavor
  violation, lepton (g \ensuremath{-} 2)$_{\mu, e}$ and electron EDM in the
  modular symmetry}},
  \href{https://doi.org/10.1007/JHEP08(2022)013}{\emph{JHEP} {\bfseries 08}
  (2022) 013} [\href{https://arxiv.org/abs/2204.12325}{{\ttfamily
  2204.12325}}].

\bibitem{deAlwis:2009fn}
S.P.~de~Alwis, \emph{{Classical and Quantum SUSY Breaking Effects in IIB Local
  Models}}, \href{https://doi.org/10.1007/JHEP03(2010)078}{\emph{JHEP}
  {\bfseries 03} (2010) 078} [\href{https://arxiv.org/abs/0912.2950}{{\ttfamily
  0912.2950}}].

\bibitem{deAlwis:2012vp}
S.P.~de~Alwis, \emph{{Constraints on LVS Compactifications of IIB String
  Theory}}, \href{https://doi.org/10.1007/JHEP05(2012)026}{\emph{JHEP}
  {\bfseries 05} (2012) 026} [\href{https://arxiv.org/abs/1202.1546}{{\ttfamily
  1202.1546}}].

\end{thebibliography}\endgroup

\end{document}